\documentclass{article}
\usepackage{spconf,amsmath,amssymb,algorithm,algpseudocode,xcolor,graphicx,cite}
\usepackage[mathscr]{eucal}
\usepackage{amsbsy}

\usepackage{bm}
\newcommand{\V}[1]{{\bm{\mathbf{\MakeLowercase{#1}}}}} % vector
\newcommand{\M}[1]{{\bm{\mathbf{\MakeUppercase{#1}}}}} % matrix
\DeclareMathOperator*{\argmin}{arg\,min}

\title{MULTIVIEW SENSING WITH UNKNOWN PERMUTATIONS: \\
AN OPTIMAL TRANSPORT APPROACH}
\name{Yanting Ma$^*$, Petros~T.~Boufounos$^*$, Hassan~Mansour$^*$, Shuchin~Aeron$^\dagger$\thanks{SA performed this work at MERL and acknowledges support by NSF CAREER award \#1553075. PB, HM, YM are exclusively funded by MERL.}}
\address{$^*$Mitsubishi Electric Research Laboratories (MERL),{\em \{yma,petrosb,mansour\}@merl.com}.\\$^\dagger$Electrical and Computer Engineering, Tufts University, {\em shuchin@ece.tufts.edu.}}

\begin{document}
\ninept
\maketitle
\begin{abstract}
In several applications, including imaging of deformable objects while in motion, simultaneous localization and mapping, and unlabeled sensing, we encounter the problem of recovering a signal that is measured subject to unknown permutations. In this paper we take a fresh look at this problem through the lens of optimal transport (OT). In particular, we recognize that in most practical applications the unknown permutations are not arbitrary but some are more likely to occur than others. We exploit this by introducing a regularization function that promotes the more likely permutations in the solution. We show that, even though the general problem is not convex, an appropriate relaxation of the resulting regularized problem allows us to exploit the well-developed machinery of OT and develop a tractable algorithm. 
\end{abstract}

\section{Introduction and Motivation}
In this paper, we explore the problem of recovering a signal measured through a linear system while undergoing a partially known permutation. The problem arises in many contexts, including coherent imaging, simultaneous localization and mapping (SLAM) and partially labeled sensing, among others. Our  motivation is coherent imaging of moving deformable targets under partially known or partially observable deformations. This has a number of applications, including radar imaging of humans in motion, and in-vivo coherent imaging of moving cells, organisms and live organs, such as beating hearts and breathing lungs.

A common feature in these applications is that, under mild assumptions, the motion can be described as a transformation---typically a permutation of  pixels with respect to a reference view---of the reflectivity of the object in front of the measurement system. We assume that the motion is partially known up to a small error, modeled as an unknown permutation from the true position to the assumed known position. In practice, motion is typically estimated using an auxiliary measurement system or a motion model\cite{tuzel2008learning,UAOKVSBP2013,AHBR2013,fischer2015surrogate, wulff2015efficient}.

This problem has strong connections to the general problem of unlabeled sensing or shuffled linear regression~\cite{unnikrishnan2015, haghighatshoar2017, unnikrishnan2018,pananjady2017, pananjady2018,hsu2017, dokmani2018, tsakiris2019}, in which the labels or indices of the data are not available during the data acquisition process. In other words, the data has undergone a completely unknown random permutation (shuffling) that also needs to be recovered during reconstruction. The motivating example for this category of problems is typically SLAM~\cite{unnikrishnan2015,unnikrishnan2018}. However, the unknown shuffling is rarely completely unknown in practice, even in the case of SLAM. Typically, there is prior knowledge of the permutation, maybe up to small errors, e.g.,~\cite{AldroubiGrochenig2001}.  

Our paper makes the following contributions:
\begin{enumerate}
    \item A formulation that explicitly regularizes the permutation to be estimated. This incorporates prior knowledge of the permutation, reducing the search space, and, in principle, making the problem easier to solve.
    \item A relaxation that allows us to use the well-developed theoretical and algorithmic machinery of optimal transport (OT)~\cite{COT_book} in obtaining a solution.
    \item A generalization of the unlabeled sensing problem, introducing an optional linear operator measuring the permuted data. This model is more appropriate for coherent imaging under motion, our motivating application.
\end{enumerate}

In the next section we introduce the problem and key assumptions. Section~\ref{sec:related_work} describes prior work on unlabeled sensing, in the context of this paper. In Section~\ref{sec:algorithm} we outline our approach to solving the problem, drawing insights from the theory and methods of OT~\cite{COT_book,Cuturi13,OTAM_book,xie2020fast}. In Section~\ref{sec:exp} we present results on synthetic data validating our approach. Section~\ref{sec:conclusion} discusses our findings and concludes.

\begin{figure}[t]
    \centering
    \includegraphics[width=0.45\textwidth]{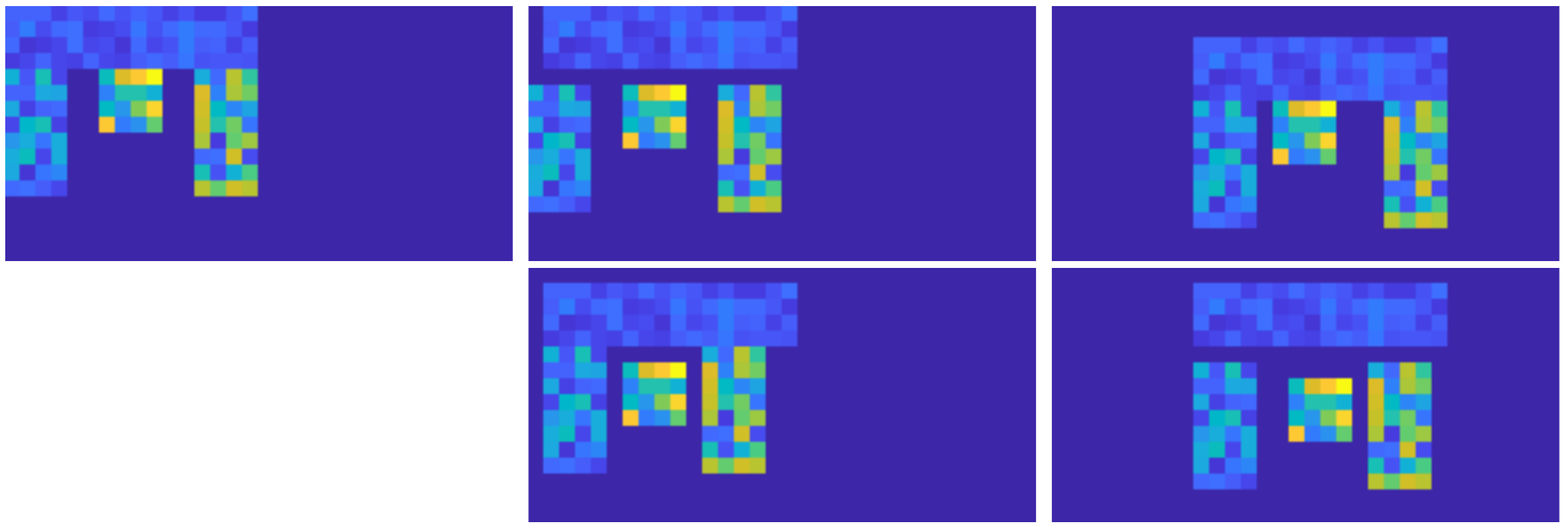}
    \caption{Illustrative example of our set up. Left: Signal $\V{x}$. Middle and Right: (top) estimates $\M{F}_i\V{x}$ of two different permutations of $\V{x}$, (bottom) actual permutations $\V{x}_i=\M{P}_i\M{F}_i\V{x}$ of $\V{x}$ observed by the acquisition system.}
    \label{fig:example}
\end{figure}

\section{Acquisition Model}
\label{sec:formulation}
We assume multiple shots $\V{y}_i$ of a reference reflectivity image $\V{x}\in\mathbb{R}^N$ are acquired through 
\begin{align}
\label{eq:measurement_model}
    \V{y}_i = \M{A}_i \M{P}_i \M{F}_i \V{x} + \V{n}_i,~~i = 1,2, ... K
\end{align} 
where,
\begin{enumerate}
    \item $\M{A}_i$ are known linear measurement operators, modeling the acquisition system,
	\item $\M{F}_i$ are known operators that partly predict the deformation of $\V{x}$, into $\V{x}_i$, 
	before it is acquired, and 
	\item $\M{P}_i$ are {\em unknown} permutation matrices modeling the uncertainty in our knowledge of $\M{F}_i$.
\end{enumerate}  
In the context of our motivating example, $\M{A}_i$ describes the coherent imaging system, such as radar, magnetic resonance, or ultrasound. The operators $\M{F}_i$ describe the object motion, e.g., of the heart or the lung, and are typically estimated using auxiliary information, such as a motion model or an alternative sensing modality; we do not explore how these are estimated in this paper. The unknown permutations $\M{P}_i$ model the correction to the estimation error of $\M{F}_i$, and may be estimated either implicitly or explicitly. Our ultimate goal is to recover $\V{x}$ from the measurements $\V{y}_i$. Estimating the $\M{P}_i$ is incidental, and we are not interested in the quality of this estimation.

Figure \ref{fig:example} provides an illustrative stylized example of our set up. The letter ``E" is observed under motion of each of its components. The top left shows the image in the reference pose. The top middle and right show the image under the estimated transformation, i.e., $\M{F}_i\V{x}$, and the bottom middle and right show the actual image $\V{x}_i$ that the acquisition system acquires through $\M{A}_i$. The assumption is that the latter is an unknown limited permutation of the former, i.e., $\V{x}_i=\M{P}_i\M{F}_i\V{x}$, as shown in the figure. The goal is to recover the reference image in the top left of the figure. 

To this end we make the following mild assumptions. 
\begin{itemize}
    \item[(a)] The support of $\V{x}$ is known;
    \item[(b)] Perturbations $\M{P}_i$ that move pixels far from their estimated position in the 2D image domain are less likely.
\end{itemize}

In the next section we will briefly review some of the related work and contrast our problem formulation. Although related, there are some salient aspects of our set up that makes it unique and that renders itself to novel solution approaches.

\section{Related work}
\label{sec:related_work}
% Our multiview sensing problem falls within the class of signal recovery problems where the target signal is measured through a linear operator that observes permuted linear transformations of the unknown signal and where the permutations are also unknown. 

% Suppose that the target signal to be recovered is $\V{x} \in \mathbb{R}^N$ and let a single view measurement be $\V{y}_i \in \mathbb{C}^M,$ where $i$ denotes the view index. Then our measurement model can be described by
% \begin{equation}
% 	\V{y}_i = \M{A} \M{P}_i \M{F}_i \V{x} + \V{n}_i,
% 	\label{eq:measurement_model}
% \end{equation}
% where $\M{A}: \mathbb{R}^N \rightarrow \mathbb{R}^M$ is the known measurement operator, $\M{P}_i$ is an unknown, but distance constrained, permutation for view $i$ that has a limited range for permuting the entries, $\M{F}_i: \mathbb{R}^N \rightarrow \mathbb{R}^N$ is a known transformation operator for view $i$, and $\V{n}_i$ is a noise vector.

Our model relates to certain signal estimation problems with unknown permutations known as \emph{unlabeled sensing} or \emph{shuffled linear regression}~\cite{unnikrishnan2015,haghighatshoar2017,unnikrishnan2018,pananjady2017,pananjady2018,hsu2017,dokmani2018,tsakiris2019}. The unlabeled sensing problem can be defined as finding a signal $\V{x}$ and a general permutation $\M{P}$, such that 
\begin{align}
\label{eq:uls}
    \V{y} = \M{P}\M{B}\V{x} + \V{n},
\end{align}
where $\M{B}$ is an $M\times N$ matrix with column rank $N,$ and $N < M$. 

These problems are instances of well-studied data association and assignment problems~\cite{Rainer_SIAM},~\cite[Ch. 2]{COT_book}. OT, a key component of our solution, is one approach to handling the 2-D assignment problem. The problem we address is an instance of an N-D (N$>2$) assignment problem~\cite{POORE20061074, Emami_2018}, which is harder to solve in general. Our approach solves smaller 2-D assignment problems, effectively, computing a \emph{barycenter}, an approach that is again inspired by and is related to the OT barycenter problem~\cite{pmlr-v32-cuturi14}. Thus we avoid computationally expensive methods that solve the N-D problem~\cite{POORE20061074, Emami_2018}.

In addition, our formulation has a key difference that necessitates our approach: we target a more general form of the unlabeled sensing problem which includes the (possibly underdetermined) linear measurement operator $\M{A}$. This precludes the aforementioned approaches. Still, if we specify $\M{A}$ as the identity operator, we solve a regularized form of the problem, incorporating additional knowledge that some permutations are more likely to occur.

Specifically, we observe multiple permuted transformations of our signal, wherein  permutations for each view are less likely to  occur the farther they diverge from the identity operator. Here, we may consider having an equivalent operator to the matrix $\M{B}$ which can be composed by stacking the multi-view transformation matrices $\M{F}_i$ into a single matrix. We also note that, we are only interested in recovering the reflectivity, not necessarily the permutation. 

The variation of the unlabeled sensing problem we consider, allows us to develop a computationally efficient algorithm, described in Section~\ref{sec:algorithm}. Experimentally, our approach performs well under our assumptions, even in the presence of noise, overcoming existing pessimistic computational results for~\eqref{eq:uls}.

\def\grad{\nabla}

\section{Recovery Using Permutation Regularization}
\label{sec:algorithm}

Considering that all permutations are not equally likely in inverting~\eqref{eq:measurement_model}, it becomes natural to consider formulations in which the permutation estimation is regularized. In particular, to estimate $\M{P}_i$ and $\V{x}$ from \eqref{eq:measurement_model}, the following is a reasonable formulation:
\begin{equation}
\min_{\M{P}_i \in \mathcal{P}, \V{x}}\, \sum_{i=1}^K \left(\frac{1}{2}\| \V{y}_i  - \M{A}_i \M{P}_i \M{F}_i \V{x} \|_2^2 + \beta R(\M{P}_i)\right),
\label{eq:formulation1a}
\end{equation}
where $\mathcal{P}$ is a set of $N\times N$ permutation matrices and $R(\M{P}_i)$ is a regularization for $\M{P}_i$, which incorporates the prior knowledge that $\M{P}_i$ is more likely to move elements to nearby pixels rather than distant ones, following by Assumption (b). 

In particular, $R(\M{P}_i)$ should penalize permutations that move pixels very far from the original position, as measured in the underlying grid of the signal. Using $l[n]$ to denote the true position of coefficient $n$ in the underlying grid $\mathcal{L}$, we assume the cost of permuting coefficient $n$ to position $n'$ to be the squared Euclidean distance in the underlying grid, $\|l[n]-l[n']\|_2^2$. For example, if the underlying grid is 2-dimensional, as in the examples in Section~\ref{sec:exp}, $l[n]$ would be the 2-dimensional position of the $n^\mathrm{th}$ coefficient of $\V{x}$. Thus, the total regularization cost of permutation matrix $\M{P}_i$ is 
\begin{equation}
R(\M{P}_i):=\sum_{n,n'=1}^{N} \| l[n]- l[n']\|_2^2\, \M{P}_i[n,n'].
\label{eq:def_R}
\end{equation}
This cost promotes permutations with small deviations from the identity, i.e., controlled errors. Of course, when appropriate for the application, other distance metrics, such as the $\ell_1$-norm can be used instead of the squared Euclidean distance. 

Solving~\eqref{eq:formulation1a} is hard in general, as it requires a combinatorial search. Thus, we consider a relaxation that allows us to use the well-established algorithmic machinery of OT~\cite{COT_book,Cuturi13,OTAM_book}. To do so, note that~\eqref{eq:formulation1a} is equivalent to
\begin{equation}
\begin{split}
&\min_{\M{P}_i \in \mathcal{P}, \V{x},\V{x}_i} \, \sum_{i=1}^K \left(\frac{1}{2} \| \V{y}_i  - \M{A}_i \V{x}_i \|_2^2 + \beta R(\M{P}_i)\right),\\
&\text{subject to }\, \V{x}_i -  \M{P}_i \M{F}_i \V{x} = 0,\quad \forall i.
\end{split}
\label{eq:formulation1b}
\end{equation}
Next, we relax the equality constraint to
\begin{equation}
\begin{split}
&\min_{\M{P}_i \in \mathcal{P}, \V{x},\V{x}_i} \, \sum_{i=1}^K \left(\frac{1}{2}\| \V{y}_i  - \M{A}_i \V{x}_i \|_2^2 + \beta R(\M{P}_i)\right),\\
&\text{subject to }\,  \| \V{x}_i -  \M{P}_i \M{F}_i \V{x} \|_2^2 \leq t,\quad \forall i.
\end{split}
\label{eq:formulation2}
\end{equation}
Writing the above in Lagrangian form, we have
\begin{equation}
\min_{\M{P}_i \in \mathcal{P}, \V{x},\V{x}_i} \sum_{i=1}^K\left(  \frac{1}{2} \| \V{y}_i  - \M{A}_i \V{x}_i \|_2^2 + \beta R(\M{P}_i) + \frac{\lambda}{2} \| \V{x}_i -  \M{P}_i \M{F}_i \V{x} \|_2^2\right).
\label{eq:formulation3}
\end{equation}
where $t$ and $\lambda$ are inversely related. Since $\M{P}_i$ is a permutation matrix, the final term in~\eqref{eq:formulation3} can be expressed as
\begin{equation}
\| \V{x}_i -  \M{P}_i \M{F}_i \V{x} \|_2^2 = \sum_{n,n'=1}^N \left( \V{x}_i[n] - (\M{F}_i\V{x})[n'] \right)^2 \M{P}[n,n'].
\label{eq:rewrite_L2}
\end{equation}

The first term in~\eqref{eq:formulation3} is independent of $\M{P}_i$ and the last two terms are linear in $\M{P}_i$. Therefore, we can incorporate~\eqref{eq:def_R} and consolidate them by defining the cost matrix $\M{C}(\V{x}_i, \M{F}_i\V{x})$ as
\begin{equation*}
\M{C}(\V{x}_i, \M{F}_i\V{x})[n,n']:=\| l[n]- l[n']\|_2^2 + \frac{\lambda}{2\beta}  \left( \V{x}_i[n] - (\M{F}_i\V{x})[n'] \right)^2 ,
\end{equation*}
and rewriting \eqref{eq:formulation3} as 
\begin{equation}
\min_{\V{x},\V{x}_i } \, \sum_{i=1}^K\left( \| \V{y}_i  - \M{A}_i \V{x}_i \|_2^2 + \beta \min_{\M{P}_i \in \mathcal{P}}\, \langle \M{C}(\V{x}_i, \M{F}_i\V{x}) , \M{P}_i \rangle \right).
\label{eq:formulation4}
\end{equation}
The minimization over $\M{P}_i$ in~\eqref{eq:formulation4} is known as the 2-D assignment problem, which is equivalent to solving a linear program~\cite{COT_book}. Specifically, this linear programming relaxation %of the 2-D assignment problem 
can be interpreted as determining an optimal probabilistic coupling, i.e., a joint distribution $\M{P}\in\Pi(\V{u},\V{v})$, where
$\Pi(\V{u},\V{v})=\{ \M{P} : 0 \leq \M{P}_{i,j} \leq 1, \M{P}\V{1} = \V{u}_i, \M{P}^T\V{1} = \V{v}_i \}$, 
between two probability distributions $\V{u}_i\in [0,1]^N$ and $\V{v}_i\in [0,1]^N$ defined
on the grid $\mathcal{L}$, where $\V{1}$ is a vector of all ones.  Then with $\V{u}_i=\V{v}_i=\V{1}/N$, \eqref{eq:formulation4} is equivalent to
\begin{equation}
\min_{\V{x},\V{x}_i }\! \sum_{i=1}^K\left( \| \V{y}_i  - \M{A}_i \V{x}_i \|_2^2 \!+\!\beta\!\! \min_{\M{P}_i \in \Pi(\V{u}_i, \V{v}_i)}\! \langle \M{C}(\V{x}_i, \M{F}_i\V{x}) , \M{P}_i \rangle \right)
\label{eq:formulation5}
\end{equation}

Next, we revisit \eqref{eq:formulation5} based on the theory of OT~\cite{COT_book,OTAM_book} and outline an efficient algorithm to solve the resulting formulation. 

\begin{algorithm}[t]
\caption{Estimate single view image $\hat{\V{x}}_i =\textsf{F}_1(\M{A}_i, \V{y}_i, \V{x})$}\label{algo:single_shot}
\textbf{Compute:} , $\V{z}_i=\M{F}_i\V{x}$ and $\V{v}_i = a(\V{z})$.
\begin{algorithmic}[1]
\For{\texttt{t = 1 to tMax}}
	\State $\V{u}_i^{t} = a(\V{x}_i^{t})$
	\State $\M{P}_i^*  = \displaystyle\argmin_{\M{P}_i\in \Pi(\V{u}_i^t, \V{v}_i)}  \langle \M{C}(\V{x}_i^t, \V{z}_i) , \M{P}_i\rangle$
	\State $\V{x}_i^{t+1} = \V{x}_i^{t} - \gamma_t \grad_{\V{x}_i}  f(\V{x},\V{x}_i^t)$
        \EndFor
\end{algorithmic}
\textbf{Output:} $\hat{\V{x}}_i = \V{x}_i^{\texttt{tMax}}$
\end{algorithm}

\begin{algorithm}[t]
\caption{Estimate prototype image $\hat{\V{x}} = \textsf{F}_2(\V{x}_1,\ldots,\V{x}_K)$}\label{algo:prototype}
\textbf{Compute:} $\V{u}_i = b(\V{x}_i)$, $i=1,\ldots,K$.
\begin{algorithmic}[1]
\For{\texttt{t = 1 to tMax}}
	\For{\texttt{i=1 to K}}
	\State $\V{v}_i^{t} = a(\M{F}_i\V{x}^{t})$
	\State $\M{P}_i^*  = \displaystyle\argmin_{\M{P}\in \Pi(\V{u}_i, \V{v}_i^t)}  \langle \M{C}(\V{x}_i, \M{F}_i\V{x}^t) , \M{P}\rangle$
	\EndFor
	\State $\V{x}^{t+1} = \V{x}^{t} - \gamma_t \sum_{i=1}^K  \grad_{\V{x}}  f(\V{x}^t,\V{x}_i)$
        \EndFor
\end{algorithmic}
\textbf{Output:} $\hat{\V{x}} = \V{x}^{\texttt{tMax}}$
\end{algorithm}

\begin{algorithm}[t]
\caption{OT Regularized Multiview Sensing}\label{algo:MM_OT}
\textbf{Input:} $\M{A}_i$, $\M{F}_i$, $\V{y}_i$, $i=1,\ldots,K$.\\
\textbf{Initialization for prototype image:} $\V{x}^0$.
\begin{algorithmic}[1]
\For{\texttt{t = 1 to tMax}}
        \For{\texttt{i = 1 to K}}       
        \State $\V{x}_i^t = \textsf{F}_1(\M{A}_i, \V{y}_i, \V{x}^{t-1})$    
        \EndFor       
        \State $\V{x}^t = \textsf{F}_2(\V{x}_1^t,\ldots,\V{x}_K^t)$
\EndFor
\end{algorithmic}
\textbf{Output:} $\hat{\V{x}} = \V{x}^{\texttt{tMax}}$
\end{algorithm}

\section{Efficient Recovery Using Optimal Transport Based Relaxation}
Our primary goal is to estimate $\V{x}$ and not necessarily the permutations. Keeping in mind the relation between OT and assignment problems \cite{Rainer_SIAM} \cite[Ch. 2]{COT_book}, we further make the following relaxations: (a) The marginals $\V{u}, \V{v}$ are assumed to be general, i.e. not necessarily uniform distributions, with possibly different supports; (b) This necessitates relaxing the the constraint on $\M{P}_i$ being permutation matrices to $\M{P}_i$ being \emph{couplings} between marginals $\V{u}, \V{v}$. In this case the estimate obtained from the inner minimization corresponds to what is referred to as a ``plan" (i.e. a joint distribution with the given marginals) in the OT literature \cite{COT_book,OTAM_book}. In other words we seek a softer coupling instead of a hard assignment. Note that when the supports are equal and the marginals are uniform, then the optimal plan and optimal assignment coincide.  Thus, by mapping our problem to an OT one, we can apply efficient algorithms developed in that literature, such as \cite{Cuturi13,xie2020fast}, to solve for the optimal plan.

\textbf{Choice of marginals}: Given a signal $\V{x}[n]$, we choose the following marginals, which seem to provide good numerical results.
Let $a: \mathbb{R}_+^{N} \to [0,1]^N$ be a function that maps reflectivity values to a probability distribution, defined as
\begin{equation}
a(\V{x})[n] := \frac{\mathbb{I}\{\V{x}[n] > T \}}{\sum_{k=1}^N \mathbb{I}\{\V{x}[k] > T\}},\quad n=1,\ldots,N,
 \label{eq:def_a}
\end{equation}
where $\mathbb{I}$ is an indicator function and $T>0$ is some predefined threshold. Thus, $a(\V{x})$ is a uniform distribution over the grid points with sufficiently large reflectivity values.
We then define the marginals as $\V{u} = a(\V{x})$, $ \V{v} = a(\M{F}_i \V{x})$ and an optimal transport distance between  $a(\V{x}_i)$ and $a(\M{F}_i \V{x})$ as
\begin{equation}
{\text{OT}}(a(\V{x}_i), a(\M{F}_i \V{x})) = \min_{\M{P}_i \in \Pi( a(\V{x}_i), a(\M{F}_i \V{x}))} \, \langle \M{C}(\V{x}_i, \M{F}_i\V{x}) , \M{P}_i \rangle
\label{eq:def_W}
\end{equation}

Given the marginals, we propose to solve for the following relaxed version of \eqref{eq:formulation4}:
\begin{equation}
\begin{split}
& \min_{\V{x},\V{x}_i }\, \sum_{i=1}^K f(\V{x},\V{x}_i),\text{ where }\\
& f(\V{x},\V{x}_i)=\| \V{y}_i  - \M{A}_i \V{x}_i \|_2^2 + \beta\, {\text{OT}}(a(\V{x}_i), a(\M{F}_i \V{x})).
\end{split}
\label{eq:formulation_final}
\end{equation}
It is worth mentioning that ${\text{OT}}(a(\V{x}_i), a(\M{F}_i \V{x}))$ may be written as a transportation-$L^p$ distance \cite{trillos2016continuum, thorpe2017} as $d_{TL^2}^2((\V{x}_i, \V{u}), (\M{F}_i\V{x}, \V{v}))$.
The particular case considered here is when the marginals are a function of the signal. While this case may fall in the general set up discussed in \cite{trillos2016continuum, thorpe2017}, it is unclear whether the theoretical results therein still hold when the marginals depend on the signal; we leave the study of theoretical properties of \eqref{eq:formulation_final} for future work.

In order to use gradient descent methods to solve \eqref{eq:formulation_final}, we need to compute the gradient of ${\text{OT}}(a(\V{x}_i), a(\M{F}_i \V{x}))$ defined in \eqref{eq:def_W} with respect to $\V{x}_i$ and $\V{x}$.
Let $\V{f}$ and $\V{g}$ be Lagrangian multipliers, the Lagrangian form of \eqref{eq:def_W} is
\begin{equation*}
\begin{split}
L(\M{P}_i,\V{f},\V{g},\V{x}_i,\V{x}) = \langle \M{C}(\V{x}_i,\M{F}_i\V{x}), \M{P}_i \rangle &+ \langle \V{f}, \M{P}_i\V{1} - a(\V{x}_i)\rangle\\
& +  \langle \V{g}, \M{P}_i^T\V{1} - a(\M{F}_i\V{x})\rangle.
\end{split}
\end{equation*}
Let $\M{P}_i^*$, $(\V{f}^*, \V{g}^*)$ be the primal and dual optima, respectively. Although they all depend on $\V{x}_i,\V{x}$, the envelope theorem \cite{SIAM_71} allows us to conveniently compute the gradient:
\begin{align*}
\grad_{\V{x}_i} \text{OT}( a(\V{x}_i),& a(\M{F}_i \V{x}))  = \grad_{\V{x}_i}  L(\M{P}_i^*,\V{f}^*,\V{g}^*,\V{x}_i,\V{x}) \\
& \overset{(a)}{=}\grad_{\V{x}_i}  \langle \M{C}(\V{x}_i,\M{F}_i\V{x}), \M{P}_i^* \rangle - \grad_{\V{x}_i} \langle \V{f}^*, a(\V{x}_i) \rangle\\
& \overset{(b)}{=} \grad_{\V{x}_i}  \langle \M{C}(\V{x}_i,\M{F}_i\V{x}), \M{P}_i^* \rangle, \text{ a.e.},
\end{align*}
where in step $(a)$, $\M{P}_i^*$ and $\V{f}^*$ are considered as constant with respect to $\V{x}_i$ (by the envelope theorem), and step $(b)$ follows by definition of $a$ in \eqref{eq:def_a}, whose gradient is zero almost everywhere with respect to Lebesgue measure on $\mathbb{R}^N$. The gradient of $\text{OT}(a(\V{x}_i), a(\M{F}_i \V{x}))$ with respect to $\V{x}$ can be computed in a similar way. Then the gradient of the cost function $f(\V{x},\V{x}_i)$ can be computed as
\begin{align*}
\grad_{\V{x}} f(\V{x},\V{x}_i) &= \lambda \M{F}_i^T \Big( a(\M{F}_i \V{x}) \odot (\M{F}_i \V{x}) - (\M{P}_i^*)^T \V{x}_i  \Big),\\
\grad_{\V{x}_i} f(\V{x},\V{x}_i)  &= \M{A}_i^T (\M{A}_i\V{x}_i - \V{y}) + \lambda  \Big( a(\V{x}_i) \odot \V{x}_i - \M{P}_i^* \M{F}_i \V{x}  \Big),
\end{align*}
where $\odot$ denotes point-wise product.

Our proposed algorithm for estimating $\V{x}$ from \eqref{eq:measurement_model} solves \eqref{eq:formulation_final} by alternating between the estimation of $\V{x}_i$ and $\V{x}$. With a fixed $\V{x}$, we estimate $\V{x}_i$ by Algorithm \ref{algo:single_shot}, and with a fixed $\V{x}_i$, we estimate $\V{x}$ by Algorithm \ref{algo:prototype}; the full algorithm is summarized in Algorithm \ref{algo:MM_OT}. The estimation of $\M{P}_i^*$ in Line~3 of Algorithm~\ref{algo:single_shot} and Line~4 of Algorithm~\ref{algo:prototype} are solved by IPOT~\cite{xie2020fast}.

\begin{figure}[t]
    \centering
    \includegraphics[width=0.45\textwidth]{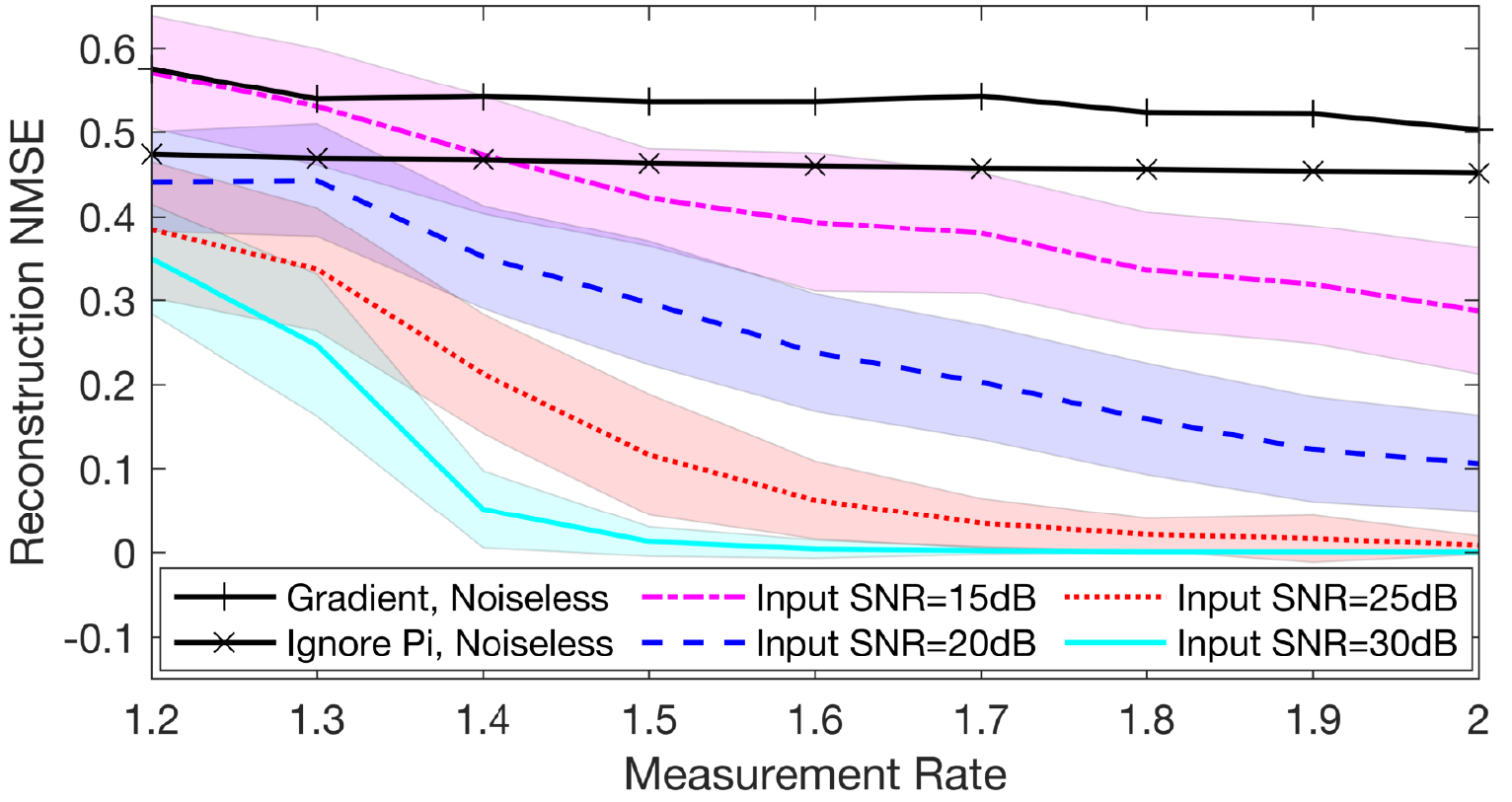}
    \caption{NMSE as a function of measurement rate at various input SNR, where the number of views is 2. The shaded area represents one standard deviation below and above the mean.}
    \label{fig:recSNR_measRate}
\end{figure}

\section{Experiments on Synthetic Data}
\label{sec:exp}

To validate our proposed algorithm, we perform two different sets of experiments on simulated data. For the first set, the number of views is fixed to be 2 and we test the algorithm at different measurement rates, defined as the measurement rate per view times the number of views, and input signal to noise ratios (SNR), defined as $\|\M{A}_i\V{x}_i\|_2^2/\|\V{y}_i - \M{A}_i\V{x}_i\|_2^2$. 

\begin{figure}[t]
    \centering
    \includegraphics[width=0.42\textwidth]{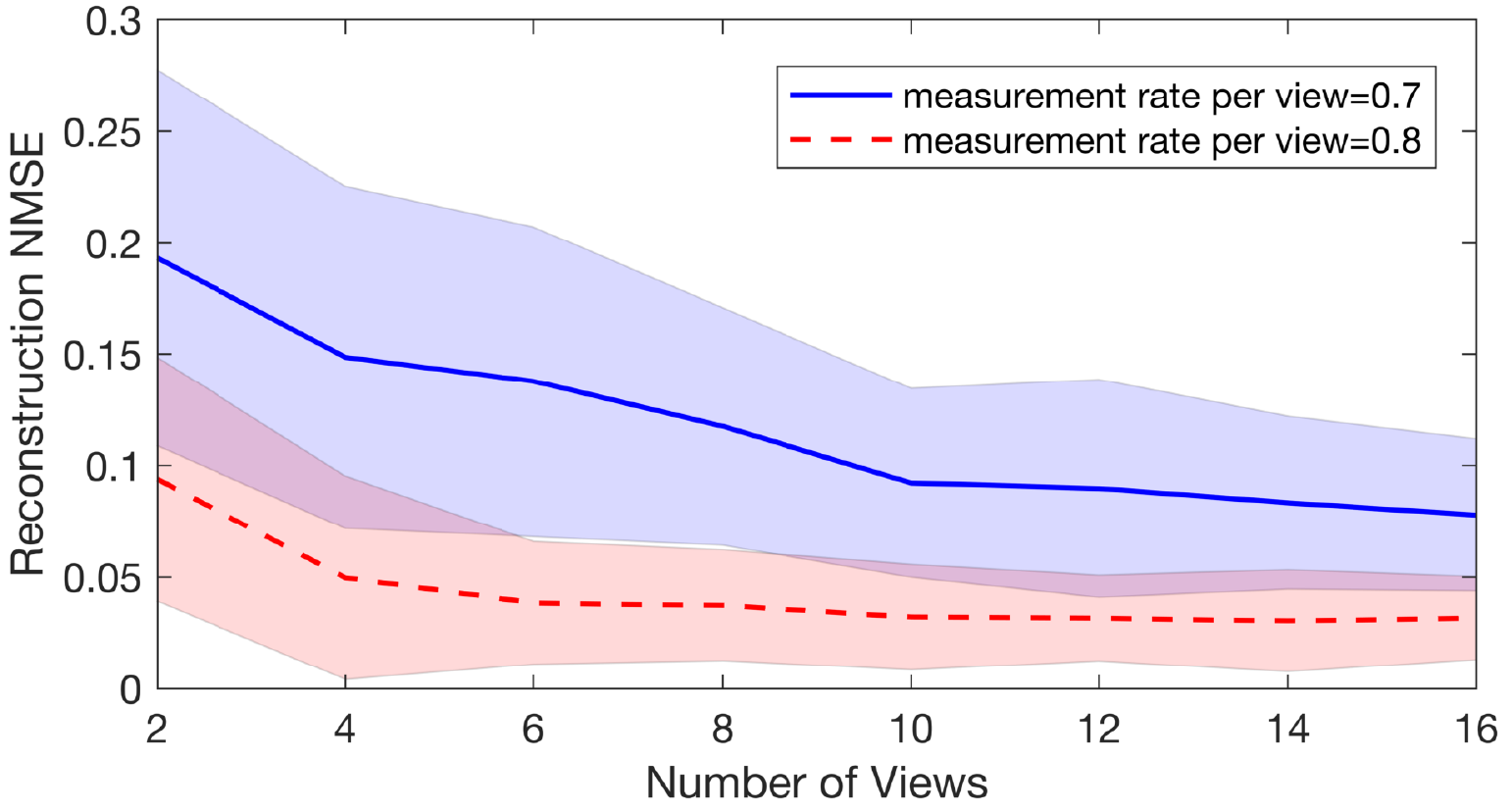}
    \caption{NMSE as a function of number of views. The measurement rate per view is 0.7 or 0.8 and the input SNR is 20dB.The shaded area represents one standard deviation below and above the mean.}
    \label{fig:recSNR_numViews}
\end{figure}

In this set of experiments, to demonstrate the effectiveness of the proposed method, we include results for two baseline methods. The first one, labeled Gradient, is a straightforward approach to solving \eqref{eq:formulation1a} directly, by alternating between estimating $\V{x}$ and $\M{P}_i$ and solving each subproblem using gradient descent. The constraint $\M{P}_i\in\mathcal{P}$ is relaxed to $\M{P}_i\in [0,1]^{N\times N}$ with an additional regularizer to promote $\M{P}_i$ having the same row and column sums as a permutation:  $\|\M{P}_i \mathbf{1} - \mathbf{1}\|_2^2 + \|\M{P}_i^T \mathbf{1} - \mathbf{1}\|_2^2$. The second baseline method, labeled Ignore $\M{P}_i$ in the figure, solves $\V{x}$ from \eqref{eq:formulation1a} assuming that $\M{P}_i$ is identity.

For the second set of experiments, the input SNR is fixed to be 20dB and the measurement rate per view is fixed to be 0.7 or 0.8. We test the algorithm at different numbers of views. A key issue we explore is that while increasing the number of views increases measurements, it also increases the number of unknown permutations. Given their poor performance of the baselines in the first set, we did not include them in this set.

In all experiments, the reference pose $\V{x}$, the estimated transformation $\M{F}_i\V{x}$, and the actual measured image $\V{x}_i$ are generated in a similar way as the example in Figure~\ref{fig:example}, except that in the second experiment, the letter ``T" instead of ``E" is used to simplify data generation. The number of pixels in each image is $N=512$. The support of the ground truth $\V{x}$ is known to the algorithm and thus so is the support size $K$. The threshold $T$ in \eqref{eq:def_a} is set to be the $K^{th}$ largest pixel value in the corresponding vector.
The measurement matrix $\M{A}_i$ has i.i.d. Gaussian entries with mean zero and variance $1/N$. Performance is measured by the normalized mean squared error (NMSE), defined as $\|\hat{\V{x}} - \V{x}\|_2^2/\|\V{x}\|_2^2$.

Figure~\ref{fig:recSNR_measRate} presents the results for the first experiment. The figure shows that the two baseline methods perform poorly even without measurement noise. For our proposed method,
the reconstruction performance improves as the measurement rate increases.
Moreover, when the input SNR is 25dB or higher, the reconstruction performance improves rather quickly.

Figure~\ref{fig:recSNR_numViews} presents the results for the second experiment. It shows that an increased number of views improves reconstruction of the reference pose, despite introducing more unknown permutations. Moreover, for a given input SNR, measurement rate, and sparsity of the reference image, performance seems to stop improving after a certain number of views; we defer comprehensive investigation of this effect to future work.

\section{Discussion and Conclusions}
\label{sec:conclusion}
Estimation of signals observed under unknown permutations is a difficult problem in general. Recognizing that in many applications some permutations are more likely than others, we introduce a regularization term which promotes certain permutations over others. By further relaxing the problem, we are able to exploit well-developed techniques in OT to provide tractable algorithms for this problem. A key component in this formulation is the judicious choice of the OT ground cost to incorporate the regularization penalty. While we present a particular choice for this cost, there are several other options, depending on the application, the study of which we defer to later publications.

\bibliographystyle{IEEEtran}
\bibliography{MM}

\end{document}